\documentclass[conference]{IEEEtran}
\IEEEoverridecommandlockouts
\usepackage{cite}
\usepackage{amsmath,amssymb,amsfonts}
\usepackage{algorithmic}
\usepackage[dvipdfmx]{graphicx}
\usepackage{textcomp}
\usepackage{xcolor}
\usepackage{multirow}
\usepackage{hyperref}
\usepackage{url}

\def\BibTeX{{\rm B\kern-.05em{\sc i\kern-.025em b}\kern-.08em
    T\kern-.1667em\lower.7ex\hbox{E}\kern-.125emX}}

\makeatletter
\def\blfootnote{\gdef\@thefnmark{}\@footnotetext}
\makeatother
\newcommand{\authormarkone}{$^*$}

\begin{document}

\title{Recommending Extract Method Refactoring\\Based on Confidence of Predicted Method Name
}

\author{}
\author{\IEEEauthorblockN{Jinto Yamanaka \authormarkone}
\IEEEauthorblockA{\textit{Graduate School of} \\ \textit{Science and Technology} \\
\textit{University of Tsukuba} \\
Tsukuba, Japan \\
yamanaka@kde.cs.tsukuba.ac.jp}
\and
\IEEEauthorblockN{Yasuhiro Hayase \authormarkone}
\IEEEauthorblockA{\textit{Faculty of} \\ \textit{Engineering, Information and Systems} \\
\textit{University of Tsukuba}\\
Tsukuba, Japan \\
hayase@cs.tsukuba.ac.jp}
\and
\IEEEauthorblockN{Toshiyuki Amagasa}
\IEEEauthorblockA{\textit{Center for Computational Sciences} \\
\textit{University of Tsukuba}\\
Tsukuba, Japan \\
amagasa@cs.tsukuba.ac.jp}
}

\maketitle
\blfootnote{\authormarkone The first two authors contributed equally to this work}

\begin{abstract}
Refactoring is an important activity that is frequently performed in software development, and among them, Extract Method is known to be one of the most frequently performed refactorings.
The existing techniques for recommending Extract Method refactoring calculate metrics from the source method and the code fragments to be extracted to order the recommendation candidates.
This paper proposes a new technique for accurately recommending Extract Method refactoring by considering whether code fragments are semantically coherent chunks that can be given clear method names, in addition to the metrics used in previous studies.
As a criterion for the semantic coherency, the proposed technique employs the probability (i.e. confidence) of the predicted method names for the code fragments output by code2seq, which is a state-of-the-art method name prediction technique.
The evaluation experiment confirmed that the proposed technique has higher correctness of recommendation than the existing techniques.
\end{abstract}


\section{Introduction}
In software development, refactoring which improves the design of the source code without changing the external behavior is an essential and frequent activity to maintain the readability and changeability of the software\cite{Fowler_1999, Mens_2004_survey}. Among them, Extract Method refactoring which extracts a part of code from an existing method as a new method is known to be one of the most frequently performed ones\cite{Murphy_2006, Chatzigeorgiou_2010}.

On the other hand, identifying where and how to refactor gets more difficult as a software product becomes larger and more complex.
To address this problem, various techniques have been proposed to support developers' refactoring activities.
As for Extract Method, there are several techniques to suggest to developers what part of which method should be extracted\cite{Tsantalis_2011_JDeodrant_LongMethod, Silva_2014_JExtract, Silva_2015_JExtract, Charalampidou_2017_SEMI, Xu_2017_GEMS}.
These existing techniques use various metrics such as complexity, coupling, and cohesion obtained from the source method and the code fragment to be extracted as a new method for recommending refactoring.

Now, we focus on the name of the newly created method in Extract Method refactoring.
Extract method refactoring is completed by giving a clear method name that expresses its role and meaning to the extracted code.
Therefore, the semantic coherence of the extracted code fragments, in the sense that the developers can give them a clear name, is one of the most influential factors in the developer's decision on whether to perform Extract Method or not.
However, the existing techniques do not take into account the semantic cohesiveness of the code to be extracted in recommending Extract Method.

This paper proposes a new technique for Extract Method recommendation that applies code2seq\cite{Alon_2019_codeseq}, a method name estimation technique, to a part of code to be extracted, and uses the confidence value of the estimated name in addition to the metrics used by existing methods.
Because the confidence value of code2seq is correlated to the degree of correspondence between the name of the method and its content, the proposed technique is expected to reflect whether the part of the code to be extracted is semantically coherent enough to be given a clear name.
In order to utilize the metrics and recommendation algorithms used by the existing methods, the proposed technique is implemented by extending GEMS\cite{Xu_2017_GEMS}, which is the state of the art of Extract Method recommendation.

This paper is organized as follows.
Section \ref{Related Work} introduces Extract Method refactoring, method name recommendation, and existing recommendation techniques as prior knowledge of this research; Section \ref{Proposed Technique} describes the proposed technique; Section \ref{Evaluation Experiment} shows the evaluation experiments; and finally, Section \ref{Conclusion} summarizes this research.

\section{Related Work}
\label{Related Work}

\subsection{Extract Method Refactoring}
Extract Method refactoring is a technique to separate a method by extracting some code from an existing method as a new method\cite{Fowler_1999}.
The purpose of this refactoring is to improve the comprehension of the program by splitting a method that is too long, or a method that implements multiple features into each one.
This refactoring consists of the following steps.

\begin{enumerate}
  \item Determine the code to be extracted from the existing methods.
  \item Create a new method by extracting the determined code into the body.
  \item Give the new method a clear name that identifies the role and behavior of the code.
  \item Replace the extracted code fragments with the call to the new method in the source method.
\end{enumerate}

\begin{figure}[tb]
  \centering
  \includegraphics[keepaspectratio, scale=0.31]{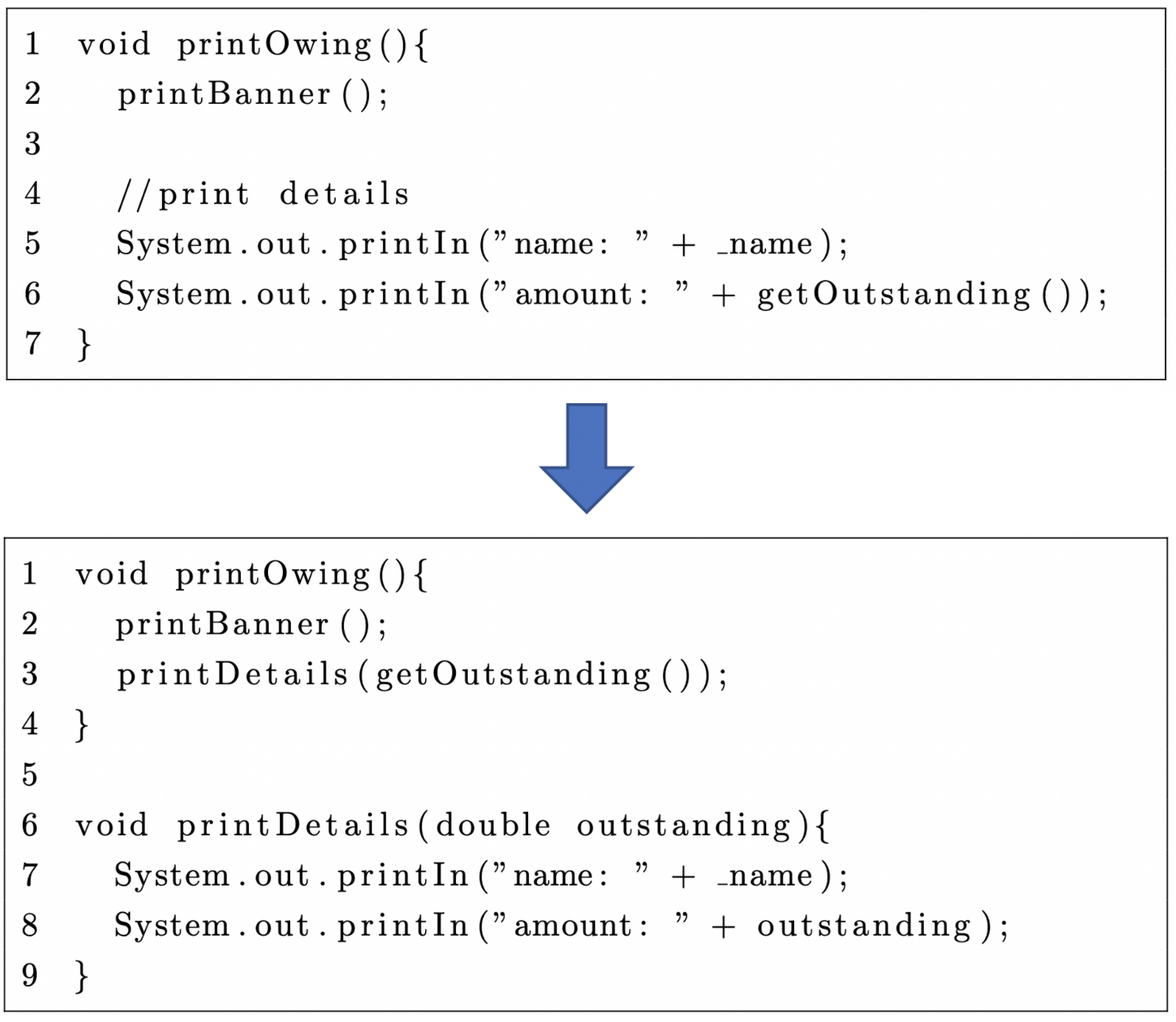}
  \caption{Example of Extract Method refactoring\cite{Fowler_1999}}
  \label{Example of Extract Method refactoring}
\end{figure}

Fig. \ref{Example of Extract Method refactoring} is an example of Extract Method refactoring shown in \cite{Fowler_1999}.
The method {\it printDetails} is newly created by extracting lines 5-6 from the method {\it printOwing} in this figure.
A statement calling {\it printDetails} is added instead of the extracted code fragment in the source method {\it printOwing}.

Extract Method is one of the most frequently applied refactorings.
Murphy et al.\cite{Murphy_2006} found that Extract Method refactoring was performed by more than 50\% of the developers surveyed among the 11 major types of refactorings performed by the Eclipse refactoring function.
Also, in the usage statistics of JDeodorant\footnote{\url{https://users.encs.concordia.ca/~nikolaos/stats.html}}, a refactoring plugin for Eclipse, this refactoring accounts for almost 50\% of all performed refactorings.

\subsection{Method Name Recommendation}
Høst et al. analyzed the relationship between the behavior of methods and the verbs used in the method names\cite{Host_2007}. They analyzed the typical behavior of methods including the verb in their names and identified the typical behavior for 40 concrete verbs. Also, Høst et al. proposed a technique that alerts the naming bugs of methods to developers and that recommends how to fix the naming bugs\cite{Host_2009}. The technique suggests a list of method names according to the semantic distance between a method name and implementation of the method.

Kashiwabara et al. proposed a technique to recommend candidate verbs for a method name so that developers can use consistent verbs for method names\cite{Kashiwabara_2014}.
They have identified four meaningful groups of rules for verb recommendation as follows: the first group of rules recommends the same verb as methods called in the method. The second group recommends verbs that are conceptually related to a certain word in the method. The third group recommends verbs related to a class definition The fourth group recommends verbs based on the Java programming idioms.

Allamanis et al. proposed how to generate method names by inputting a sequence of tokens appearing in the source code into a neural convolutional attentional model that includes a convolutional network within the attention mechanism itself\cite{Allamanis_2016}. Also, they presented deep learning models for recommending method names by modeling the code’s graph structure and learning program representations over those graphs\cite{Allamanis_2018}.

Code2seq proposed by Alon et al.\cite{Alon_2019_codeseq} is a technique that generates a distributed representation of a method using the Abstract Syntax Tree created from the source code of the method body and predicts the method name to input this representation into the NMT model from the seq2seq paradigm.
This technique is an extension of code2vec\cite{Alon_2019_codevec} proposed by them and receives as input the source code of the method whose name is to be predicted and returns the candidates of the predicted method name in order of the probability of correct prediction (i.e. confidence).
The predicted method names are output as a sequence of words.
As far as we know, code2seq is currently the most accurate method name prediction technique.

\begin{figure*}[t]
  \centering
  \includegraphics[keepaspectratio, scale=0.305]{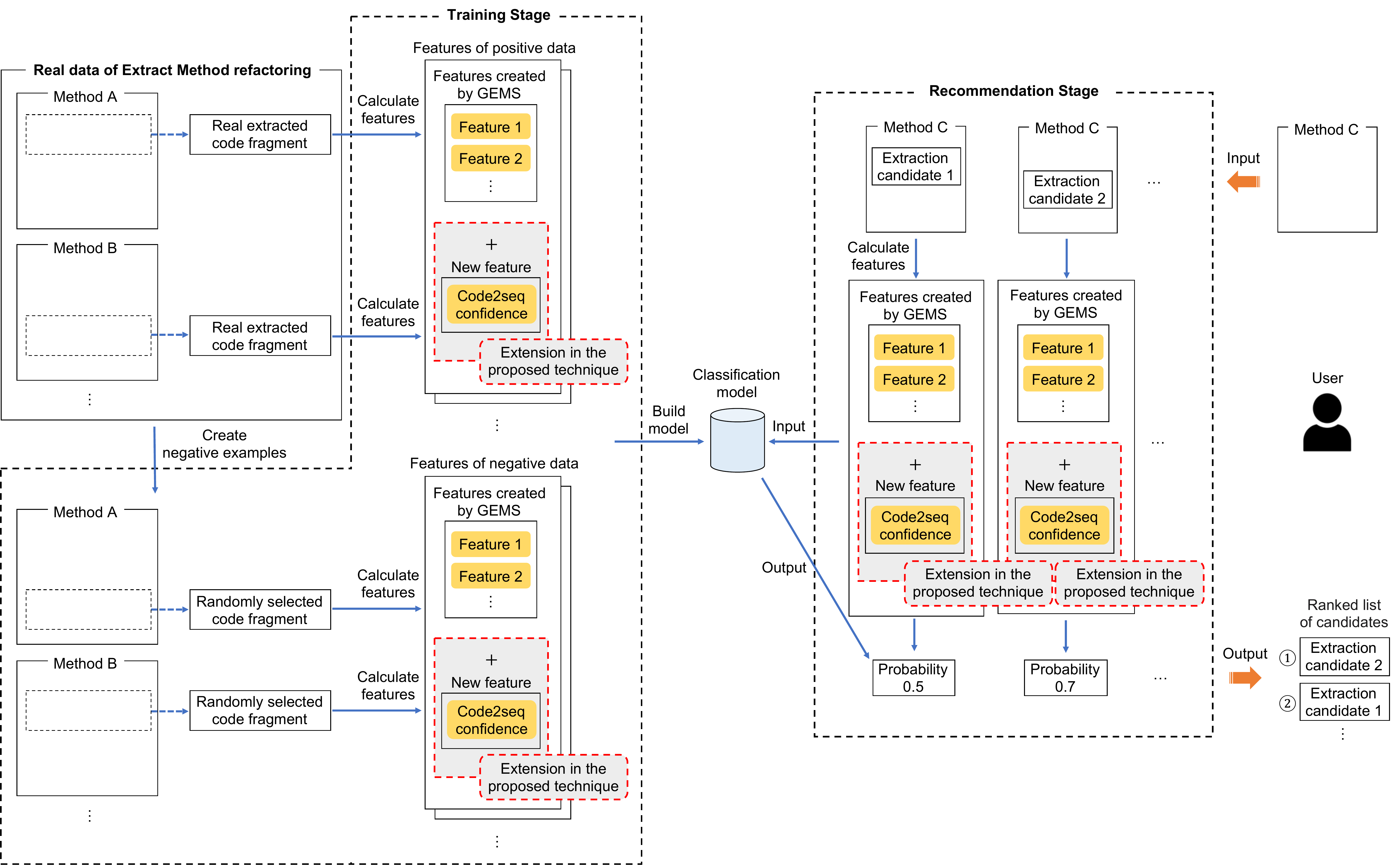}
  \caption{Overview of the proposed technique}
  \label{Overview of the proposed technique}
\end{figure*}

\subsection{Refactoring Support Techniques}
Various techniques have been proposed to support developers' refactoring activities.
Terra et al.\cite{Terra_2018_MoveMethod} and Kurbatova et al.\cite{Kurbatova_2020_MoveMethod} proposed a technique for recommending Move Method refactoring, and Bavota et al.\cite{Bavota_2011_ExtractClas, Bavota_2014_ExtractClass} proposed one to automatically identify Extract Class refactoring opportunities.
Besides, a detailed review and comparison of existing techniques has been compiled by Pecorelli et al.\cite{Pecorelli_2019} and Baqais et al.\cite{Baqais_2020}.

As for Extract Method refactoring, there are several techniques to recommend to developers what part of which method should be extracted.
JDeodorant proposed by Tsantalis et al.\cite{Tsantalis_2009_JDeodrant_FeatureEnvy, Tsantalis_2010_JDeodrant_TypeStateChecking, Tsantalis_2011_JDeodrant_LongMethod, Fokaefs_2012_JDeodrant_GodClass, Tsantalis_2017_JDeodrant_DuplicatedCode} is a method that provides multiple types of refactoring support.
Regarding Extract Method refactoring\cite{Tsantalis_2011_JDeodrant_LongMethod}, it relies on the concept of program slicing which is a technique for extracting only the source code that affects arbitrary variables in the program from the original program.
Specifically, the source code containing all the statements that change the value of a certain variable or the state of a certain object given in the body of the method is selected as the target for extraction.

Silva et al.\cite{Silva_2014_JExtract, Silva_2015_JExtract} proposed JExtract, which determines the need for extraction by obtaining metrics from a set of variables, types, and packages used in a method.
First, the technique generates all the code fragments that can be extracted from the method as extraction candidates.
The code fragments are created by eliminating the combinations that cause compilation errors when extracted after dividing the source code of the method into units of nested structures called blocks and obtaining all the combinations of consecutive statements in the same block.
Next, sets of variables, types, and packages are created from the source method and each candidate code fragment, respectively.
Then, the Kulczynski set similarity coefficient is obtained using the sets and the candidates with small similarities to the source method are recommended as extraction targets.

SEMI proposed by Charalampidou et al.\cite{Charalampidou_2017_SEMI} uses the method level $\text{LCOM}_\text{2}$ which measures the lack of class cohesion to determine the extraction target.
The degree of cohesion is an index showing how much the functionalities of the source code are aggregated.
SEMI creates candidate code fragments for extraction according to the unique algorithm and recommends extraction targets from among them according to the Benefit that is calculated using $\text{LCOM}_\text{2}$ obtained from the source method and each extraction candidate.

Xu et al.\cite{Xu_2017_GEMS} proposed a method called GEMS.
They created a classification model that consists of 48 different metrics as features to identify the extraction targets, because the metrics used in existing techniques only consider specific program elements, and that the approach that relies on specific metrics is not practical since actual refactoring doesn't only aim to improve the metrics.
GEMS uses the same extraction algorithm as JExtract to create code fragments for the extraction candidates and creates features from each candidate.
Then, it classifies each candidate as to whether it should be extracted or not based on the model and recommends those that are classified as extraction targets in order of the highest prediction probability.
As far as we know, GEMS has shown the best recommendation results among the existing Extract Method recommendation techniques.

\section{Proposed Technique}
\label{Proposed Technique}
This paper proposes a technique to improve the recommendation correctness by considering whether candidate code fragments are semantically coherent chunks that can be given clear method names code fragments for extraction.
As a criterion for the semantic coherency, the proposed technique employs the probability (i.e. confidence) of the predicted method names for the code fragments output by code2seq \cite{Alon_2019_codeseq}, which is a state-of-the-art method name prediction technique.
The proposed method is based on the idea that since extract method refactoring is completed by giving a clear name to the newly created method that expresses its role and behavior, it is important for the refactoring decision whether the code fragment to be extracted has a semantic coherence that can be given such a name.
The implementation of the proposed technique is based on GEMS, which is the state of the art of Extract Method recommendation techniques, and uses the confidence of the method name output by code2seq as an explanatory variable in addition to the features used by GEMS.
Note that code2seq and GEMS used in the implementation are the ones taken from the GitHub\footnote{\url{https://github.com}} repository\cite{Code2seq_Github} and the ones distributed as plugins, respectively.

Fig. \ref{Overview of the proposed technique} shows the overview of the proposed technique.
The overall structure of the proposed method is similar to that of GEMS and consists of two stages. The first stage is the training stage, which builds a classification model to determine whether a code fragment is a good candidate for extraction or not, and is executed prior to the second stage. The second stage is the recommendation stage, which uses the constructed model to recommend a candidate of Extract Method, and is executed on request of the user.

\subsection{Training Stage}
\label{Training Stage}
The training stage takes as input a set of examples of Extract Method refactorings, and outputs a statistical classification model that determines whether a given part of a method is suitable to be extracted as a method.
The framework of this stage is almost the same as that of the same stage in GEMS, with the only difference being the addition of the confidence of code2seq as a feature.
The example of an extracted method is a set of the entire source code of the original method and what part of the code was extracted as a new method.

Due to the requirements of the classification algorithm used by the technique, parts of the method that should not be extracted (i.e., negative examples) are required to build the classification model, so negative examples are generated from the (positive) examples of the Extract Method.
Specifically, from the source code of the positive example, a part of the code that is different from the actual extraction is randomly selected, and the pair of the source code and the part is used as the negative example.

The features are then extracted from the positive and negative examples, and the set of features is fed to a learning algorithm to build the classification model.
Totally 49 types of features are given to the learning algorithm, including the 48 types of features used in GEMS (Table \ref{Structural features generated by GEMS} and Table \ref{Functional features generated by GEMS}) and the confidence values obtained by applying code2seq to the part of the source code that are candidates for extraction.


\subsection{Recommendation Stage}
The recommendation stage is executed in response to a request from a user, and outputs a ranked list of the parts of the method specified by the user that are considered suitable for extraction as a method.
First, the stage lists all possible parts of the user-specified method that can be extracted as a method within the constraints of the programming language. Next, for each pair that combines the specified method with each part, the 49 features including the confidence of code2seq are calculated.
The features are then input to the classification model constructed in the training stage to obtain the probability that the extract method represented by each pair is likely to be performed.
Finally, a list of pairs with probabilities higher than a certain threshold is created and provided to the user in order of the probability.


\subsection{Features for classification}
\begin{figure}[t]
  \centering
  \includegraphics[keepaspectratio, scale=0.40]{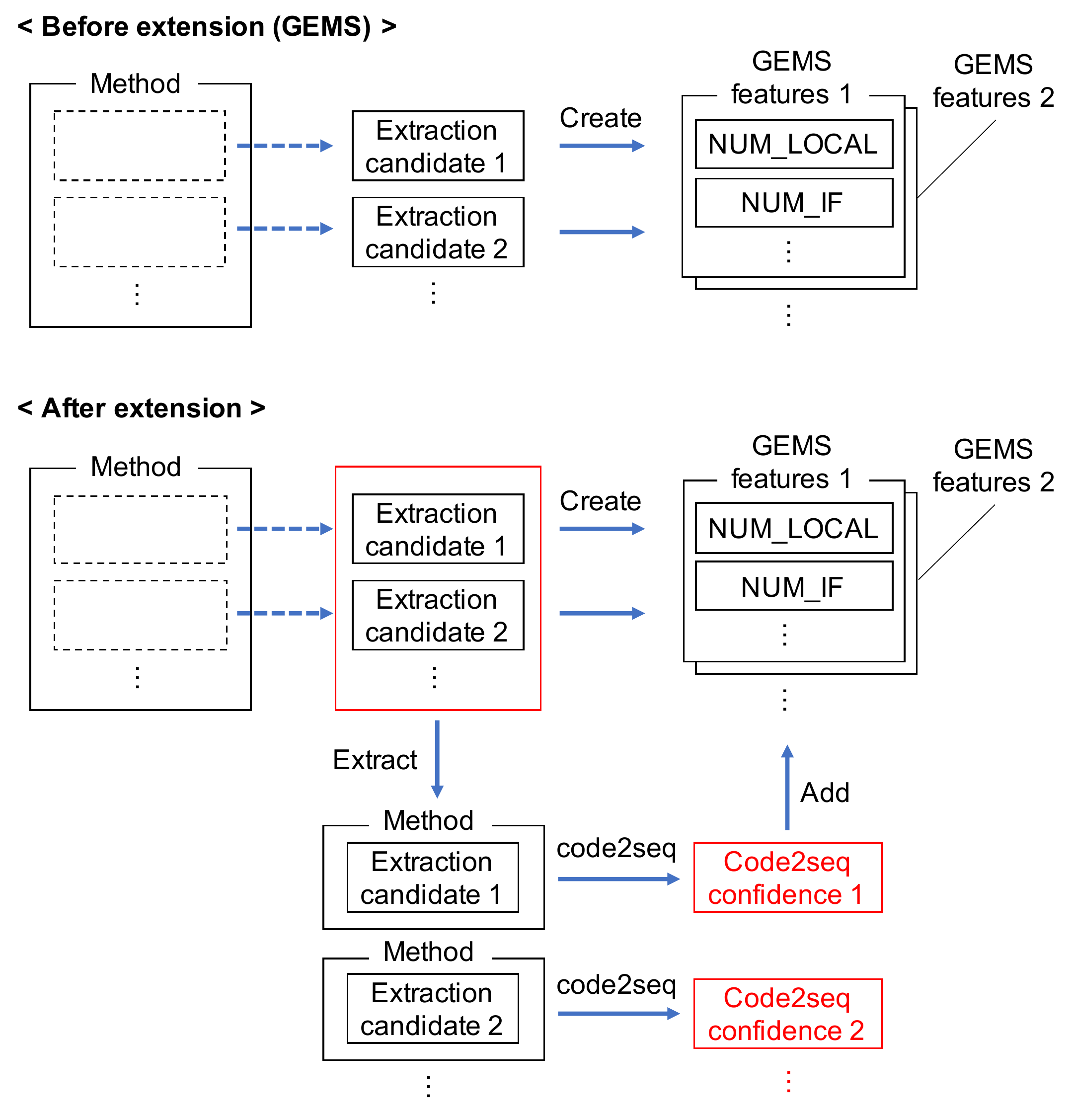}
  \caption{How to add the confidence to GEMS features}
  \label{How to add the confidence to GEMS features}
\end{figure}

The confidence that the code2seq calculates when predicting the method names of code fragments is used as an indicator of whether or not it is easy to give clear method names to candidate code fragments for extraction, and this value is added to existing features to extend GEMS.
An overview of the GEMS extension is shown in Fig. \ref{How to add the confidence to GEMS features}.
When GEMS is applied to a certain method, all the code fragments that can be extracted are generated as extraction candidates, and 48 different metrics are created as features from the pairs of the source method and each code fragment.
Then, each code fragment is converted into the form of a method by the method extraction function of Eclipse JDT and input to code2seq because it can only receive source code in method form.
After that, the confidence of the method name predicted by code2seq at the top is added to the features created by GEMS, for each code fragment.

The features used in GEMS are shown in Table \ref{Structural features generated by GEMS} and Table \ref{Functional features generated by GEMS}.
These are divided into two categories: structural features and functional features. The structural features include 28 types, and the functional features include 20 types.
The structural features shown in Table \ref{Structural features generated by GEMS} are mainly created from each of the code fragments that are extraction candidates and the remaining code obtained by removing the fragments of extraction candidates from the source method.
The functional features shown in Table \ref{Functional features generated by GEMS} are created in the following way.
First, for each program element, the element whose ratio of the number used in the candidate code fragment to the number used in the source method is 1st or 1st - 2nd is identified.
This ratio is the feature corresponding to the {\it Usage rate in extraction candidates} in the table.
Next, the ratio of the lines of code in which the identified element is used to the lines of code in the candidate code fragment is calculated.
This ratio is the feature corresponding to the {\it Dedication to elements with high usage} in the table.

\begin{table*}[t]
  \centering
  \caption{Structural features generated by GEMS}
  \renewcommand{\arraystretch}{1.3}
  \begin{tabular}{|l|l|l|} \hline
      \multicolumn{1}{|c|}{\textbf{Description of the features}} & \multicolumn{1}{|c|}{\textbf{The candidate code fragment}} & \multicolumn{1}{|c|}{\textbf{The remaining code}} \\ \hline \hline
      Lines of code & LOC\_EXTRACTED\_METHOD & CON\_LOC \\ \hline
      Number of local variables defined & NUM\_LOCAL & CON\_LOCAL \\ \hline
      Whether literals are defined or not & NUM\_LITERAL & CON\_LITERAL \\ \hline
      Number of method invocations & NUM\_INVOCATION & CON\_INVOCATION \\ \hline
      Number of if statement & NUM\_IF & CON\_IF \\ \hline
      Number of conditional operators & NUM\_CONDITIONAL & CON\_CONDITIONAL \\ \hline
      Number of switch statement & NUM\_SWITCH & CON\_SWITCH \\ \hline
      Number of variables accessed& NUM\_VAR\_AC & CON\_VAR\_ACC \\ \hline
      Number of types accessed & NUM\_TYPE\_AC & CON\_TYPE\_ACC \\ \hline
      Number of fields accessed & NUM\_FIELD\_AC & CON\_FIELD\_ACC \\ \hline
      Number of assignments & NUM\_ASSIGN & CON\_ASSIGN \\ \hline
      Number of typed elements & NUM\_TYPED\_ELE & CON\_TYPED\_ELE \\ \hline
      Number of packages to reference & NUM\_PACKAGE & CON\_PACKAGE \\ \hline
      Number of assert statement & CON\_ASSERT & n/a \\ \hline
      Ratio of LOC in the candidate code fragment to LOC in the source method & \multicolumn{2}{|l|}{RATIO\_LOC} \\ \hline
  \end{tabular}
  \label{Structural features generated by GEMS}
\end{table*}

\begin{table*}[t]
  \centering
  \renewcommand{\arraystretch}{1.3}
  \caption{Functional features generated by GEMS}
  \begin{tabular}{|l|l|l|} \hline
      \multicolumn{1}{|c|}{\textbf{The program element}} & \multicolumn{1}{|c|}{\textbf{Usage rate in extraction candidates (1st, 2nd)}} & \multicolumn{1}{|c|}{\textbf{Dedication to elements with high usage (1st, 2nd)}} \\ \hline \hline

      \multirow{2}{*}{Local variable} & RATIO\_VARIABLE\_ACCESS & VARAC\_COHESION \\
      & RATIO\_VARIABLE\_ACCESS2 & VARAC\_COHESION2 \\ \hline

      \multirow{2}{*}{Field} & RATIO\_FIELD\_ACCESS & FIELD\_COHESION \\
      & RATIO\_FIELD\_ACCESS2 & FIELD\_COHESION2 \\ \hline

      Method & RATIO\_INVOCATION & INVOCATION\_COHESION\\ \hline

      \multirow{2}{*}{Type} & RATIO\_TYPE\_ACCESS & TYPEAC\_COHESION \\
      & RATIO\_TYPE\_ACCESS2 & TYPEAC\_COHESION2 \\ \hline

      Typed element & RATIO\_TYPED\_ELE & TYPEDELE\_COHESION \\ \hline

      \multirow{2}{*}{Package} & RATIO\_PACKAGE & PACKAGE\_COHESION \\
      & RATIO\_PACKAGE2 & PACKAGE\_COHESION2 \\ \hline
  \end{tabular}
  \label{Functional features generated by GEMS}
\end{table*}

\section{Evaluation Experiment}
\label{Evaluation Experiment}
An evaluation experiment is conducted to confirm whether the introduction of the code2seq confidence in the proposed method contributes to the improvement of the correctness of the Extract Method recommendation.The baseline for the evaluation is GEMS, which is the state-of-the-art Extract Method recommendation technique and also the basis for the implementation of the proposed method.
The experiment shows the improvement of the correctness against the baseline and the contribution of the confidence to the improvement.



\subsection{Experimental Design}
\label{Experiment Design}
To evaluate the improvement of recommendation correctness and each feature contributes, the proposed technique is compared with GEMS as a baseline.
Using a prediction model with the confidence and the prediction model of GEMS, the recommendation results for the test data are compared using three metrics. Also, the importance of the features is examined for each model.

As in the previous studies\cite{Charalampidou_2017_SEMI, Xu_2017_GEMS}, we use the top-5 recommendations for each method of test data to evaluate the recommendation performance with three metrics: precision, recall, and F-measure.
It should be noted that all the recommended extraction candidates are classified as should be extracted.
The precision is the ratio of candidates with correct recommendation results to the total number of recommended extraction candidates.
The recall is the percentage of targets that are actually recommended out of all refactoring targets in test data.
The F-measure is the weighted average of the precision and recall, and is calculated by the following formula:

\[
  F-measure = \frac{2 \times precision \times recall}{(precision + recall)}
\]

Like previous studies\cite{Charalampidou_2017_SEMI, Xu_2017_GEMS}, the value called tolerance is introduced as a permissible limit on variation in lines of code when the correctness of the recommendation is evaluated.
The three tolerance patterns to be used are 1\%, 2\%, and 3\%. The number of the extraction source method lines is multiplied by the tolerance, and the value rounded up to the nearest whole number is used as the tolerance line.
For example, if a method with 50 lines is given and the tolerance is 3\%, the tolerance line is 2, since 50 * 0.03 = 1.5.
In this case, the extraction candidates with an error of ±2 lines from the correct recommendation are also considered to be correct targets.

Also, the Gini importance\cite{Breiman_1984} calculated by the Gradient boosting classifier of scikit-learn is used when comparing the importance of features.
The higher the Gini importance of a feature, the more important the feature is considered to be.

\subsection{Setup}
\subsubsection{Prediction Model}
\label{Prediction Model}
A prediction model with the confidence by code2seq added to GEMS features is created using machine learning.
The real Extract Method refactoring data created by Silva et al.\cite{Silva_2016} from the Java project available on GitHub is used to create the training data.
The real data is also used in the study of GEMS\cite{Xu_2017_GEMS} and consists of the methods that have undergone Extract Method refactoring and the code fragments extracted from those methods.
Using this real data, training data is created from the features of the positive and negative examples as shown in Section \ref{Training Stage}.
Since the case where only one extraction candidate as a positive example is created from a method is included, a total of 479 pieces of training data consisting of 244 positive examples and 235 negative examples were created from 244 methods as a result.
The training data has 49 dimensions.
Gradient boosting classifier\footnote{\url{https://scikit-learn.org/stable/modules/generated/sklearn.ensemble.GradientBoostingClassifier.html}} implemented in scikit-learn\cite{Pedregosa_2011_scikit-learn}, the Python machine learning library, is used as the machine learning algorithm similar to the study of GEMS.
Parameter tuning is performed using Optuna\cite{Akiba_2019_optuna} that is a hyperparameter optimization framework to determine the values of the hyperparameters in building the model.
5-fold cross-validation is performed on 479 training data which is the total of positive and negative examples, and the hyperparameters are searched so that the average of the five F-measure is as high as possible.

\subsubsection{Test Data}
The test data uses the methods that Extract Method refactoring should be executed identified from five Java projects: SelfPlanner, WikiDev, JHotDraw, Junit, and MyWebMarket.
The projects are considered to be quality open-source software, and 155 code fragments to be extracted have been identified from 130 methods across these five projects.
Among all data, the data from SelfPlanner and WikiDev were created by Tsantalis et al.\cite{Tsantalis_2011_JDeodrant_LongMethod} and used to evaluate the performance of JDeodorant.
The data from JHotDraw, Junit, and MyWebMarket were created by Silva et al.\cite{Silva_2014_JExtract} and used to evaluate the performance of JExtract.
Charalampidou et al.\cite{Charalampidou_2017_SEMI} and Xu et al.\cite{Xu_2017_GEMS} have also evaluated SEMI and GEMS using the data sets from these five projects, which are currently considered to be the most suitable data sets for evaluating the work of Extract Method refactoring recommendation.

\begin{table}[tb]
  \centering
  \caption{Comparison of correctness}
  \renewcommand{\arraystretch}{1.3}
  \begin{tabular}{|l|l||r|r|r|} \hline
      \multicolumn{1}{|c|}{\textbf{Metrics}} & \multicolumn{1}{|c||}{\textbf{Tolerance}} & \multicolumn{1}{|c|}{\textbf{GEMS}} & \multicolumn{1}{|c|}{\textbf{GEMS + Conf}} \\ \hline \hline
      \multicolumn{1}{|l|}{\multirow{4}{*}{Precision}} & None & 0.08757 & 0.09574 \\ 
      & 1\% & 0.21191 & 0.21986 \\ 
      & 2\% & 0.21366 & 0.22518 \\ 
      & 3\% & 0.22067 & 0.23227 \\ \hline
      \multicolumn{1}{|l|}{\multirow{4}{*}{Recall}} & None & 0.32258 & 0.34839 \\ 
      & 1\% & 0.45806 & 0.52258 \\ 
      & 2\% & 0.46452 & 0.52903 \\ 
      & 3\% & 0.48387 & 0.54839 \\ \hline
      \multirow{4}{*}{F-measure} & None & 0.13774 & 0.15021 \\ 
      & 1\% & 0.28977 & 0.30950 \\ 
      & 2\% & 0.29269 & 0.31590 \\ 
      & 3\% & 0.30310 & 0.32632 \\ \hline
  \end{tabular}
  \label{Comparison of correctness}
\end{table}

\subsection{Results}
\subsubsection{Comparison of Correctness}
\label{Section Comparison of Correctness}
The results of the comparison of the three metrics are shown in TABLE \ref{Comparison of correctness}.
In the table, a model labeled {\it GEMS} is created from GEMS features only, while a model labeled {\it GEMS + Conf} is created by adding confidence from code2seq to GEMS features.
The total number of the top-5 recommendations for the 130 methods is 571 using the model without confidence and 564 using the model with confidence.

From TABLE \ref{Comparison of correctness}, it can be seen that the precision, recall, and F-measure increase for all the tolerances when the confidence is used compared to the case where it is not used.
The F-measure increases by about 1.25 for no tolerance, 1.97 for 1\%, and 2.32 for 2\% - 3\%.
Then, the increase in the recall is larger than that of the precision, which is about 2.58 for no tolerance and about 6.45 for 1\% - 3\%.

In this paper, the model is built so that the number of positive and negative examples of the training data is almost equal, following the previous research, but in reality, most of the extraction candidates obtained from the method are considered to be classified as results that should not be extracted.
Therefore, the ratio of positive examples to negative examples while model training is not considered to be realistic.
So, it is necessary to note that the reliability of the precision is considered to be lower than that of recall.

\begin{table}[t]
  \centering
  \caption{Feature importance of GEMS}
  \renewcommand{\arraystretch}{1.3}
  \begin{tabular}{|r|l|r|} \hline
      \multicolumn{1}{|c|}{\textbf{Rank}} & \multicolumn{1}{|c|}{\textbf{Feature}} & \multicolumn{1}{|c|}{\textbf{Importance }} \\ \hline \hline
      1 & TYPEDELE\_COHESION & 0.15175 \\
      2 & NUM\_TYPED\_ELE & 0.12510 \\
      3 & RATIO\_LOC & 0.10388 \\
      4 & INVOCATION\_COHESION & 0.09432 \\
      5 & NUM\_PACKAGE & 0.09407 \\
      6 & CON\_LOC & 0.05041 \\
      7 & NUM\_TYPE\_AC & 0.04943 \\
      8 & VARAC\_COHESION & 0.04165 \\
      9 & CON\_TYPED\_ELE & 0.03155 \\
      10 & CON\_PACKAGE & 0.03067 \\ \hline
  \end{tabular}
  \label{Feature importance of GEMS}
\end{table}

\begin{table}[t]
  \centering
  \caption{Feature importance of the proposed technique}
  \renewcommand{\arraystretch}{1.3}
  \begin{tabular}{|r|l|r|} \hline
      \multicolumn{1}{|c|}{\textbf{Rank}} & \multicolumn{1}{|c|}{\textbf{Feature}} & \multicolumn{1}{|c|}{\textbf{Importance}} \\ \hline \hline
      1 & \textbf{CODE2SEQ\_CONFIDENCE} & \textbf{0.30011} \\
      2 & TYPEDELE\_COHESION & 0.15584 \\
      3 & INVOCATION\_COHESION & 0.08658 \\
      4 & NUM\_PACKAGE & 0.07611 \\
      5 & RATIO\_LOC & 0.05153 \\
      6 & NUM\_INVOCATION & 0.04172 \\
      7 & NUM\_VAR\_AC & 0.03058 \\
      8 & NUM\_TYPE\_AC & 0.02505 \\
      9 & RATIO\_TYPE\_ACCESS & 0.02409 \\
      10 & LOC\_EXTRACTED\_METHOD & 0.02188 \\ \hline
  \end{tabular}
  \label{Feature importance of the proposed technique}
\end{table}

\begin{table*}[t]
  \centering
  \caption{Comparison of the confidence between positive and negative data}
  \renewcommand{\arraystretch}{1.3}
  \begin{tabular}{|l||r|r|r|r|} \hline
      \multicolumn{1}{|c||}{\textbf{Label}} & \multicolumn{1}{|c|}{\textbf{Maximum}} & \multicolumn{1}{|c|}{\textbf{Minimum}} & \multicolumn{1}{|c|}{\textbf{Mean}} & \multicolumn{1}{|c|}{\textbf{Median}} \\ \hline \hline
      Positive & 0.67956 & $5.9918\times10^{-15}$ & $6.3695\times10^{-2}$ & $3.6578\times10^{-3}$ \\
      Negative & 0.76879 & $3.2844\times10^{-18}$ & $1.1233\times10^{-2}$ & $4.9674\times10^{-7}$ \\ \hline
  \end{tabular}
  \label{Comparison of the confidence between positive and negative data}
\end{table*}

\subsubsection{Comparison of Feature Importance}
\label{Comparison of Feature Importance}
The top-10 importance of the features in each model are compared.
The importance of the models without confidence is shown in TABLE \ref{Feature importance of GEMS}, and the importance of the models with confidence is shown in TABLE \ref{Feature importance of the proposed technique}.
In TABLE \ref{Feature importance of the proposed technique}, CODE2SEQ\_CONFIDENCE represents the confidence by code2seq added in the proposed technique.

From TABLE \ref{Feature importance of GEMS}, when confidence is not used in the model, the most important feature is TYPEDELE\_COHESION, with a score of 0.15175.
The difference in score between this feature and the second most important feature, NUM\_TYPED\_ELE, is about 2.67.
On the other hand, TABLE \ref{Feature importance of the proposed technique} shows that when confidence is used in the model, CODE2SEQ\_CONFIDENCE is the feature with the highest importance.
The score for this importance is 0.30011, which is about 14.84 higher than the highest score in TABLE \ref{Feature importance of GEMS}, and the difference in score with TYPEDELE\_COHESION, which is the second most important, is about 14.43.

\subsection{Discussion}
From Section \ref{Section Comparison of Correctness}, it is found that the recommendation correctness of Extract Method refactoring can be improved by adding the confidence since the F-measure increases for all the tolerance.
Also, since the increase in the recall is larger than that in the precision, it can be considered that the confidence has a significant effect in increasing the coverage of the refactoring target to be extracted in the recommendation.
Furthermore, it can be said that the confidence makes a significant contribution to the prediction among all the features from the results in Section \ref{Comparison of Feature Importance}.

Code2seq has the nature of predicting the method name by acquiring semantic information about the source code from the syntactic structure of the method.
The reason why the confidence of code2seq is excellent in identifying the target of Extract Method refactoring is considered to be that this nature allowed us to determine whether the candidate code fragments have a semantic coherence suitable for refactoring that is easy to explain their roles.
To examine whether there is a clear difference in the value of the confidence between the code fragments that are real examples for refactoring and other code fragments, the maximum, minimum, mean, and median confidence for the extracted code fragments of the positive and negative examples created from the real data in Section \ref{Prediction Model} are compared.
The results of the comparison are shown in TABLE \ref{Comparison of the confidence between positive and negative data}.
The table shows that the minimum, mean, and median confidence of the positive examples is higher than those of the negative examples, suggesting that the confidence tends to be higher for code fragments with semantic coherence that should be refactored.

\section{Conclusion}
\label{Conclusion}
This paper has proposed a technique to improve the recommendation correctness of Extract Method refactoring by using the confidence value of predicted names by code2seq, a method name prediction technique, for newly created methods by refactoring.
The proposed technique employs the metrics of GEMS, which has the highest correctness among the existing extract method recommendation techniques, in addition to the confidence values, and the implementation of the proposed technique is also based on GEMS.
The evaluation experiments comparing the proposed technique with GEMS confirmed the high preciseness of the proposed technique and also revealed that the confidence value contributes significantly to the estimation.

As a future work, an evaluation experiment with a larger data set is considered to be essential.
On the other hand, the correctness can be improved by using the confidence values of the second-ranked method names as features in addition to the top-ranked one.





\bibliography{ISSRE}
\bibliographystyle{unsrt}

\end{document}